\documentclass[twocolumn, letter]{jpsj3arxiv}

\usepackage{bm}
\usepackage{mathrsfs}
\usepackage{exscale}
\usepackage{color}
\usepackage{graphicx}

\newcommand{\AC}{AC}
\newcommand{\ACs}{ACs}
\newcommand{\QC}{QC}
\newcommand{\QCs}{QCs}
\newcommand{\agy}{AGY}
\newcommand{\agyI}{AGY$({\rm I})$}
\newcommand{\agyII}{AGY$({\rm II})$}
\newcommand{\aay}{AAY}
\newcommand{\agyIN}{Au$_{64.0}$Ge$_{22.0}$Yb$_{14.0}$}
\newcommand{\agyIIN}{Au$_{63.5}$Ge$_{20.5}$Yb$_{16.0}$}

\newcommand{\degc}{$^{\circ}$C}

\title{Superconductivity of Au-Ge-Yb Approximants with Tsai-type Clusters}

\author{
Kazuhiko~\text{Deguchi}$^{1}$\thanks{E-mail: deguchi@edu3.phys.nagoya-u.ac.jp}, 
Mika~\text{Nakayama}$^{1}$,
Shuya~\text{Matsukawa}$^{1}$,
Keiichiro~\text{Imura}$^{1}$,
Katsumasa~\text{Tanaka}$^{2}$,
Tsutomu~\text{Ishimasa}$^{2}$, and 
Noriaki~K.~\text{Sato}$^{1}$}

\inst{$^1$Department of Physics, Graduate School of Science, Nagoya University, Nagoya 464-8602, Japan\\
$^2$Division of Applied Physics, Graduate School of Engineering, Hokkaido University, Sapporo 060-8628, Japan}

\abst{
We report the emergence of bulk superconductivity in \agyIN\ and \agyIIN\ below 0.68 and 0.36 K, respectively.
This is the first observation of superconductivity in Tsai-type crystalline approximants of quasicrystals.
The Tsai-type cluster center is occupied by Au and Ge ions in the former approximant, and by an Yb ion in the latter.
For magnetism, the latter system shows a larger magnetization than the former. To explain this observation, we propose a model that the cluster-center Yb ion is magnetic.
The relationship between the magnetism and the superconductivity is also discussed.
}

\begin{document}
\maketitle


Quasicrystals (\QCs ) have been classified as the third solid because they possess long-range, quasi-periodic structures with diffraction symmetries forbidden for crystals. 
Owing to the considerable progress since the discovery of \QCs\ in resolving their geometric structure~\cite{Schechtman1984,Tsai2000,Takakura2007}, QCs are nowadays considered as a type of crystal.
For their electronic structure, on the other hand,
no long-range magnetic ordering has been observed 
although there are a few reports on superconductivity~\cite{Wong1987,Wagner1988}.
For a periodic approximant crystal (\AC), a phase whose composition is close to that of the \QC\ and whose unit cell has atomic decorations similar to those of the \QC , there are some reports on ferromagnetic or antiferromagnetic orderings.
However, superconductivity has not been discovered thus far to the best of our knowledge.

Recently, new types of magnetic \QC\ and \AC\ has been discovered: the Au-Al-Yb (\aay ) \QC\ exhibits novel quantum critical behavior as observed in Yb-based heavy-fermion materials with intermediate Yb valence~\cite{Deguchi2012}, while the \aay\ \AC\ shows heavy-Fermi-liquid behavior.
Since the diverging behavior of magnetic susceptibility as $T \rightarrow 0$ was only observed in the \QC , the quantum critical state may correspond to an electronic state unique to the \QCs , i.e., a critical state that is neither extended nor localized.

In the course of our research on the above novel phenomena, we learned from our review of the literature that the Au-Ge-Yb (\agy ) system belongs to the 1/1 \AC\ of a Tsai-type icosahedral \QC . 
According to Lin and Corbett~\cite{Lin2010}, the \agy\ system has two types of crystal structure: one contains 14 at\% Yb and the other 16 at\% Yb atoms.
Almost four non-Yb (i.e., Au and Ge) atoms occupy the center of the Tsai-type cluster in the former compound as in the case of \aay ~\cite{Ishimasa2011}, while, in the latter one, there is only a ``rattling'' Yb atom at the center of the cluster~\cite{Gebresenbut2013}. 
Hereafter, the former and latter \ACs\ are referred to as \agyI\ and \agyII, respectively. 

In this Letter, we report on our low-temperature experiments on the electrical resistivity, magnetization, ac magnetic susceptibility, and specific heat of the \agyI\ and \agyII\ \ACs .
We observe that both of the \ACs\ show superconductivity with transition temperatures $T_{\rm c}$ of $0.68$ K for \agyI\ and $0.36$ K for \agyII.
We further observe that the magnetization is much larger in the \agyII\ \AC\ than in the \agyI\ \AC .
To explain this observation, we propose a model in which the cluster-center Yb ion in \agyII\ is magnetic.
We also discuss the relationship between magnetism and superconductivity in these new systems.

Two types of polycrystalline sample, Au$_{86-x}$Ge$_{x}$Yb$_{14}$ ($20 \le x \le 24$) and Au$_{84-y}$Ge$_{y}$Yb$_{16}$ ($18 \le y \le 22$), were synthesized with the starting materials 4N (99.99\% pure)-Au, 5N-Ge, and 3N-Yb.
(Note that the composition is nominal throughout the paper.)
For both alloys, the starting materials were put in an alumina crucible, sealed in an evacuated quartz tube, and heated to 1000 \degc.
Then, the crucible was cooled to 800 \degc. After it was subsequently cooled slowly to 450 \degc,
the crucible was rapidly quenched in water for the synthesis of \agyI, while it was slowly cooled in the furnace for the synthesis of \agyII.

Structure analysis was carried out by a powder X-ray diffraction technique using Cu $K\alpha$-radiation; details of the 
experiments are described elsewhere~\cite{Ishimasa2011}.
The dc magnetization was measured using a commercial SQUID magnetometer in the temperature range between 1.8 and 300 K, and at magnetic fields of up to 70 kOe. 
Four-terminal resistivity measurements were performed using ac methods.
Ac magnetic susceptibility was measured using a driving ac magnetic field of 0.1 Oe at a frequency of 100.3 Hz. 
The specific heat measurement was performed by a conventional quasi-adiabatic heat-pulse method. 
The electrical resistivity, ac magnetic susceptibility, and specific heat measurements with zero dc magnetic field were carried out in a $^3$He cryostat down to 0.25 K, and the resistivity and ac magnetic susceptibility measurements at magnetic fields of up to 80 kOe were carried out in a dilution refrigerator down to 0.08 K.

\begin{figure}[t]
\begin{center}
\includegraphics[clip,width=1.00\columnwidth]{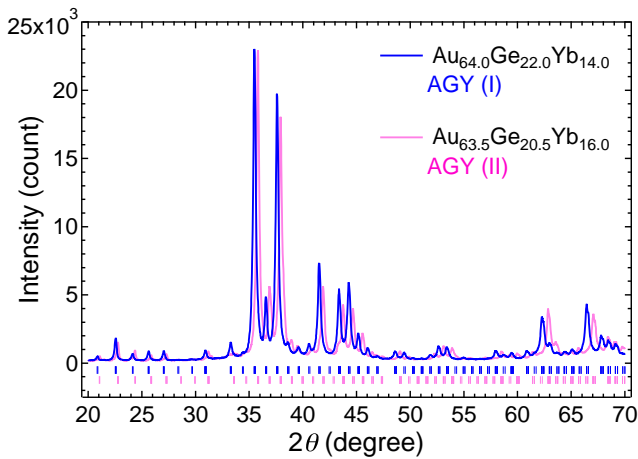}
\end{center}
\caption{(Color online) Powder X-ray diffraction patterns of \agyIN\ [\agyI ] and \agyIIN\ [\agyII ]. 
Note that the spectrum slightly shifts from each other due to the difference in the lattice parameter. 
The peak positions of \agyI\ and \agyII\ are denoted by bars in the upper and lower regions, respectively. 
}
\label{fig:X-ray}
\end{figure}
\begin{figure*}[t]
\begin{center}
\includegraphics[clip,width=2.00\columnwidth]{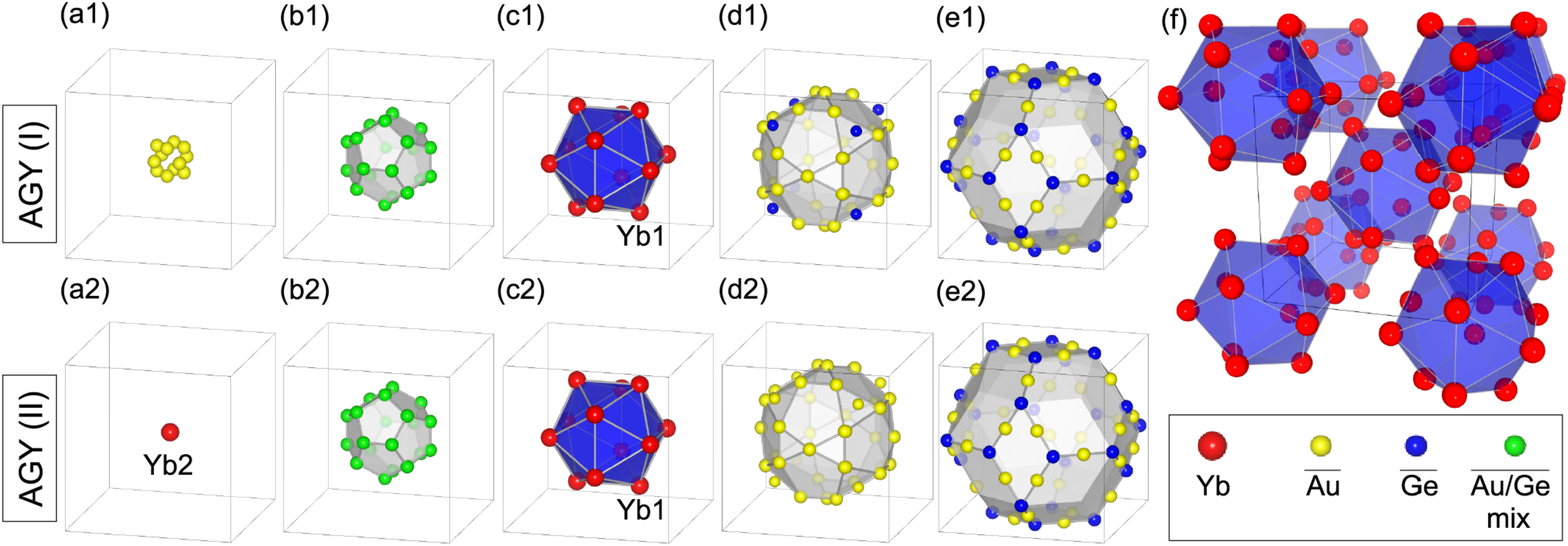}
\end{center}
\caption{(Color online) Concentric shell structure of Tsai-type cluster. (a1)-(e1) are for \agyI , and  (a2)-(e2) are for \agyII . The square frame is a unit cell. The sites including Au and Ge are classified into three groups: $\overline{\rm Au}$ (more than 90\% Au), $\overline{\rm Ge}$ (more than 90\% Ge), and mixed $\overline{\rm Au/Ge}$ (other). (a1) Orientationally disordered tetrahedron with $\overline{\rm Au}$ atoms. (a2) Single Yb ion at the center of the cluster (Yb2 site). (b1),(b2) Second shell of dodecahedron composed of mixed $\overline{\rm Au/Ge}$ atoms. (c1),(c2) Third shell of icosahedron with Yb ions on the vertex (Yb1 sites). (d1),(d2) Fourth shell of icosidodecahedron with $\overline{\rm Au}$ atoms. The atom centered at [$1/4$ $1/4$ $1/4$] is $\overline{\rm Ge}$ or $\overline{\rm Au}$. (e1),(e2) Fifth shell of triacontahedron with $\overline{\rm Au}$ and $\overline{\rm Ge}$ atoms. (f) Body-centered cubic arrangement of
the icosahedron in the Au-Ge-Yb \AC .}
\label{fig:Crystal}
\end{figure*}
Figure~\ref{fig:X-ray} shows the powder X-ray diffraction patterns of \agyIN\ and \agyIIN , both of which show a body-centered cubic structure with lattice parameters of $1.4724(2)$ and $1.4605(4)$ nm, respectively. 
Note that the alloy with a higher Yb concentration has a smaller lattice parameter, as pointed out in Ref.~\citen{Lin2010}. 

The Rietveld structure analysis of the former successfully converged and indicated that the sample with an optimal composition $x = 64.0$ is of single phase and its cluster-center tetrahedron is orientationally disordered~\cite{Deguchi2014}.
For the latter, on the other hand, the predominant phase has an optimal composition of $y = 63.5$ and the sample contains a small amount of an unidentified phase.
The inclusion of the secondary phase prevented the complete Rietveld refinement.
In the present study, we assume following Lin and Corbett that, there is a single Yb atom at the cluster center in the latter system~\cite{Lin2010, Gebresenbut2013}.
As a result, \agyI\ and \agyII\ denote \agyIN\ and \agyIIN , respectively.

Figure~\ref{fig:Crystal} shows two structure models,
both of which are composed of the Tsai-type cluster, i.e., concentric shells of a dodecahedron, an icosahedron, and an icosidodecahedron.
Note that the ``Yb1'' site at the vertex of the icosahedron is exclusively occupied by the Yb atom.
The structure model of the \agyI\ \AC\ is similar to that of the \aay\ \AC ; the cluster center is occupied by Au atoms with a probability of approximately 0.27 (Ref.~\citen{Deguchi2014}), which is close to $1/3$ in the case of a randomly oriented tetrahedron.
In contrast, in the structure model of the \agyII\ \AC , the cluster center is occupied by a single Yb atom (``Yb2'' site).

\begin{figure}[t]
\begin{center}
\includegraphics[clip,width=0.95\columnwidth]{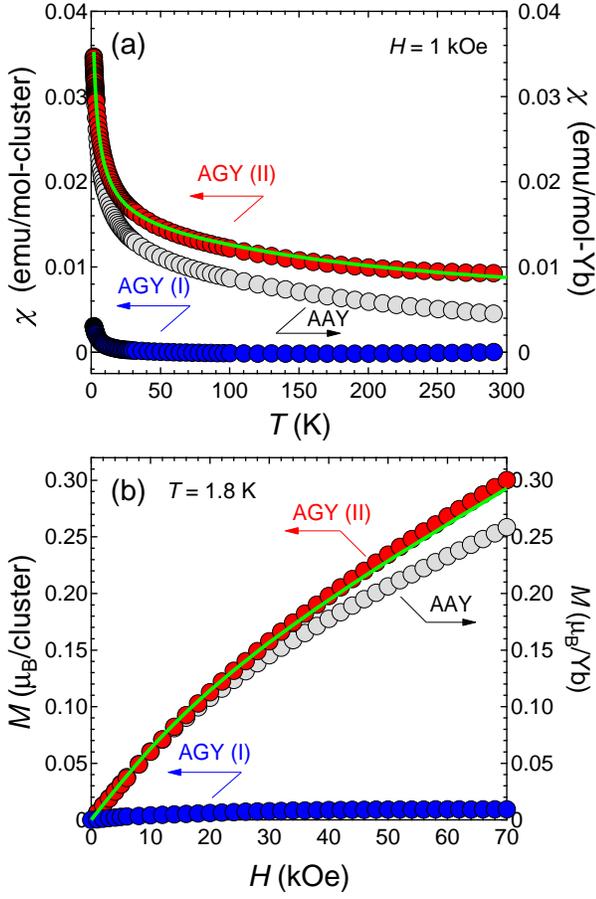}
\end{center}
\caption{(Color online) (a) Dc magnetic susceptibilities $\chi (T)$ of the \agyI, \agyII, and \aay\ \ACs\ measured at $H = 1$ kOe. 
(b) Magnetization curves $M (H)$ for the \agyI, \agyII, and \aay\ \ACs\ at $T = 1.8$ K.
Solid lines in (a) and (b) indicate calculated results; see text for details.
}
\label{fig:Mag}
\end{figure}

Figure~\ref{fig:Mag}(a) shows the temperature dependences of the magnetic susceptibilities of the \agyI\ and \agyII\ \ACs. 
For comparison,
we also plot the susceptibility data of the \aay\ \AC~\cite{Deguchi2012}.
For the analysis described below, the magnetic susceptibilities of \agyI\ and \agyII\ \ACs\ are plotted in unit per cluster (see the left vertical axis), including 12 and 13 Yb atoms, respectively, while that of the \aay\ \AC\ is given in unit per Yb ion (the right vertical axis).
The \agyI\ \AC\ shows an almost $T$-independent diamagnetic susceptibility $\chi = -1.6 \times 10^{-4}$ emu/mol-cluster at high temperatures, and shows a Curie-like rise in $\chi(T)$ in the lowest-temperature region measured.
The latter rise seems extrinsic because the low-temperature magnetization $M(H)$ saturates at a low magnetic field; as shown in Fig.~\ref{fig:Mag}(b), $M(H)$ is only $9.6 \times 10^{-3} \mu_{\rm B}$/cluster at $H = 50$ kOe.
It is reasonable to assume that the $T$-constant diamagnetism arises as a result of the cancellation between the paramagnetic contribution of conduction electrons and the diamagnetic contribution of ion cores.
Since the latter susceptibility is estimated to be $\sim -1.8\times 10^{-3}$ emu/mol-cluster from the ion core susceptibility given in the literature~\cite{SelwoodTEXT}, the Pauli susceptibility is evaluated as $\sim 1.6\times 10^{-3}$ emu/mol-cluster.  
Using a measured electronic-specific-heat coefficient $\gamma = 109$ mJ/K$^2$mol-cluster,
we obtain a Wilson ratio $R_{\rm W} = {\pi}^{2}k_{\rm B}^{2}{\chi}/3{\mu}_{\rm B}^{2}{\gamma} \sim 1$,
which is expected for a noncorrelated electron system.
As a result, the observed magnetism of the \agyI\ \AC\ 
can be understood by assuming that all the Yb ions (i.e., the Yb1-site ions at the vertex of the icosahedron) are in the nonmagnetic Yb$^{2+}$ state~\cite{Matsukawa2014}.

In contrast, the magnetic susceptibility of the \agyII\ \AC\ (containing both the Yb1- and Yb2-site ions) behaves like that of the \aay\ \AC .
Assuming that ${\chi}(T) = {\chi}_{0} + {\chi}_{4f}(T)$ [${\chi}_{0}$ and ${\chi}_{4f}(T)$ are $T$-independent and $T$-dependent contributions to the susceptibility, respectively], we find that, for $T > 80$ K, ${\chi}_{4f}(T)$ follows the Curie-Weiss law with an effective Bohr magneton $\mu_{\rm eff} = 3.5{\mu}_{\rm B}$/cluster and a Weiss temperature $T_{\theta} = 104$ K. 
Note that this effective magnetic moment per cluster is close to 3.8 $\mu_{\rm B}$/Yb of the \aay\ \AC .
Since the constant term $\chi_{0} = 4.8 \times 10^{-3}$ emu/mol-cluster is also considered as additive contributions of conduction electrons and ion cores, the Pauli paramagnetism of the \agyII\ \AC\ is estimated to be $\sim 6.8 \times 10^{-3}$ emu/mol-cluster from the ion core diamagnetism of $\sim -2.0\times 10^{-3}$ emu/mol-cluster. 
Assuming $\gamma \sim 251$ mJ/K$^2$mol-cluster at high temperatures, we obtain a Wilson ratio $R_{\rm W} \sim 2$, which suggests the possibility that the 4$f$ electrons would contribute to the itinerant electron magnetism.

Let us make a comparison between the \agyII\ \AC\ and the \aay\ \AC .
The solid line in Fig.~\ref{fig:Mag}(a) indicates that  ${\chi}_{4f}(T) = a {\chi}_{_{\rm AAY}}(T)$, where $a$ is an adjustable parameter and found to be $0.9$ and ${\chi}_{_{\rm AAY}}(T)$ is a measured susceptibility (per mole of Yb ion) of the \aay\ \AC . 
We find good coincidence between the solid curve and the data points, suggesting that Yb ions in the \agyII\ \AC\ are similar in nature to those in the \aay\ \AC.

Figure~\ref{fig:Mag}(b) shows the magnetization curves for the \agyI , \agyII , and \aay\ \ACs .
We ascribe the small magnetization of the \agyI\ \AC\ to a paramagnetic impurity effect.
The solid line indicates that $M(H) = {\chi}_{0}H + M_{4f}(H)$, where $M_{4f}(H) = a M_{\rm AAY}(H)$ [$a=0.9$ and $M_{\rm AAY}(H)$ is a measured magnetization of the \aay\ \AC\ per Yb ion].
Again, we find good agreement between the solid curve and the data points.
Taking account of the results (i) that the magnetization per cluster (containing 12 Yb1-site ions and 1 Yb2-site ion) in the \agyII\ \AC\ is similar in magnitude to that per Yb ion in the \aay\ \AC\ and (ii) that 12 Yb1-site ions are all nonmagnetic in the \agyI\ \AC , 
we propose a model showing that, in the \agyII\ \AC , the Yb1-site Yb ions are all nonmagnetic and the Yb2-site Yb ion behaves like the Yb ions in the \aay\ \AC . 
This implies that the \agyII\ \AC\ has a heterogeneous Yb valence.

\begin{figure}[t]
\begin{center}
\includegraphics[clip,width=0.85\columnwidth]{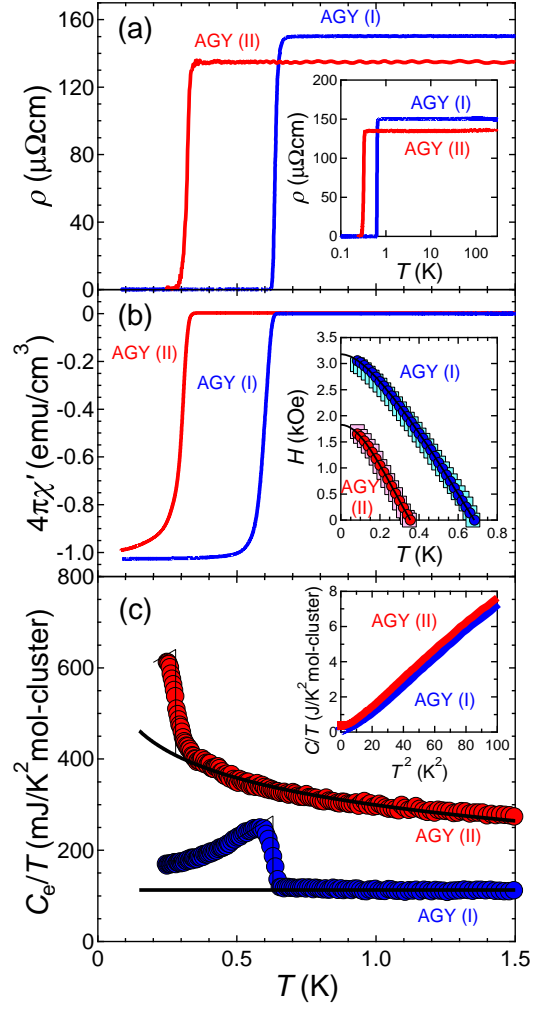}
\end{center}
\caption{(Color online) (a) Electrical resistivities $\rho (T)$ of the \agyI\ and \agyII\ \ACs . 
The inset shows the electrical resistivity on a logarithmic abscissa for $0.1 < T <300$ K. 
(b) Real parts of the ac magnetic susceptibilities $\chi' (T)$ of the \agyI\ and \agyII\ \ACs .
 The absolute value was calibrated with the superconductivity of Pb ($T_{\rm c} = 7.2$ K).
The inset shows the upper critical field $H_{\rm c2}(T)$ of the \agyI\ and \agyII\ \ACs\ deduced from the ac magnetic susceptibility (open squares) and resistivity (open circles). 
Solid lines are a guide for the eyes.
(c) Electronic specific heat divided by temperature $C_{\rm e}/T$ of the \agyI\ and \agyII\ \ACs . The low-$T$ behavior in the \agyI\ \AC\ is likely due to the nuclear contribution. The inset shows the specific heat divided by temperature $C/T$  versus $T^{2}$ for the \agyI\ and \agyII\ \ACs . }
\label{fig:SC}
\end{figure}
Figure~\ref{fig:SC}(a) shows the temperature dependences of the electrical resistivities $\rho (T)$ of the \agyI\ and \agyII\ \ACs . 
The $T$-independent feature with a large residual resistivity [see inset of Fig.~\ref{fig:SC}(a)] is characteristic of \ACs\ as well as \QCs\ with Tsai-type clusters. 
The electrical resistivities of both materials sharply drop to zero, indicating the emergence of superconductivity. 
By defining the transition temperature $T_{\rm c}$ as the midpoint of the resistivity drop, we obtain $T_{\rm c} = 0.68$ and $0.36$ K for the \agyI\ and \agyII\ \ACs , respectively. 

Figure~\ref{fig:SC}(b) shows the temperature dependences of the real parts of the ac magnetic susceptibilities $\chi' (T)$ of the \agyI\ and \agyII\ \ACs\ with zero external dc magnetic field. 
Below $T_{\rm c}$, there emerges a clear diamagnetic signal due to the superconducting shielding effect. 
When defining $T_{\rm c}$ as the onset of diamagnetism, we find that the $T_{\rm c}$ values deduced from the susceptibility and resistivity coincide with each other.

Detailed studies of the field-dependent resistivity and ac susceptibility allow us to determine the upper critical field $H_{\rm c2}(T)$; see the inset of Fig.~\ref{fig:SC}(b).
Just below $T_{\rm c}$, we observe a linear $T$ dependence with gradients of $dH_{\rm c2}/dT=5.84$ and $6.75$ kOe/K for the \agyI\ and \agyII\ \ACs , respectively.
The orbital critical field at zero temperature is estimated using this slope via the Werthamer-Helfand-Hohenberg formula $H_{\rm c2}^{\rm orb}(0) = -0.693T_{\rm c}dH_{\rm c2}/dT$, and the results are summarized in Table~\ref{tab:parameters}, together with the $H_{\rm c2}(0)$ values obtained by extrapolating the data to zero temperature. 
We find that $H_{\rm c2}^{\rm orb}(0)$ is close to $H_{\rm c2}(0)$ for both materials, suggesting that the orbital depairing mechanism dominantly contributes to the $H_{\rm c2}$ of these \ACs .

To check if the superconductivity is of bulk origin, we measured the specific heat $C(T)$.
In the normal state, $C(T)$ can be well fitted using the conventional formula $C(T) = \gamma T + \beta T^3$ ($2 < T < 5$ K) with $\gamma=109$ and $251$ mJ/K$^2$mol-cluster, and $\beta = 40.8$ and $41.1$ mJ/K$^4$mol-cluster for the \agyI\ and \agyII\ \ACs , respectively.
Using $\beta = 12{\pi}^{4}NR/5\Theta_{\rm D}^{3}$ ($R$ is the gas constant and $N$ is the number of atoms), we estimate the Debye temperature to be $\Theta_{\rm D}=160$ and 157 K for the \agyI\ and \agyII\ \ACs , respectively.
Then, by subtracting the phonon contribution of $C(T)$, we evaluate the electronic specific heat $C_{\rm e}(T)$.
In Fig.~\ref{fig:SC}(c), we plot $C_{\rm e}/T$ as a function of temperature.
A clear jump is observed at $T_{\rm c}$: the thin solid line indicates the $T$ dependence calculated using entropy balance.
The jump height is $\Delta C_{\rm e}/T_{\rm c}=138$ and $218$ mJ/K$^2$mol-cluster and therefore $\Delta C_{\rm e}/C_{\rm e}(T_{\rm c})$ is $1.26$ and $0.54$ for the \agyI\ and \agyII\ \ACs , respectively.
These values indicate the bulk superconductivity and suggest that, if the 4$f$ electrons would be itinerant, then they are involved in Cooper pair formation.

In the normal state, the \agyII\ \AC\ shows a logarithmic increase in $C_{\rm e}/T$ at low temperatures in contrast to the \agyI\ \AC : the solid line in Fig.~\ref{fig:SC}(c) indicates a fitted result using the expression $C_{\rm e}/T \sim -(S^{*}/T^{*}){\rm ln}(T/T^{*})$ with $T^{*} = 33$ K and $S^{*} = 2860 = 0.5R{\rm ln}2 $ mJ/Kmol-cluster.
Such a logarithmic $T$ dependence was observed in UPt$_3$~\cite{Stewart1984}, which allows us to suggest the possibility of an unconventional pairing mechanism related to the magnetism.

The superconducting parameters of these materials are summarized in Table~\ref{tab:parameters}.
They are evaluated using the knowledge of $T_{\rm c}$, $H_{\rm c2}(T)$ and $\Delta C_{\rm e}/T_{\rm c}$ as follows:  
The GL parameter $\kappa$ at $T = T_{\rm c}$ is deduced from the relation $4\pi \times 1.16 \times (2\kappa^{2}-1) \Delta C_{\rm e}/T_{\rm c} = (dH_{\rm c2}/dT)^{2}$, and we assume that $\kappa (0) \simeq \kappa$ from $H_{\rm c2}(0) \simeq H_{\rm c2}^{\rm orb}(0)$.
Note that the large $\kappa$ values exceeding 20 indicate that both materials are type II superconductors.
The coherence length $\xi(0)$ is estimated from $H_{\rm c2}(0) = \Phi_{0}/2\pi\xi(0)^{2}$, where $\Phi_0$ is a flux quantum.
The penetration depth is estimated from the relation $\lambda(0) = \kappa(0) \xi(0)$.
The thermodynamic critical field is evaluated from $H_{\rm c}(0) = H_{\rm c2}(0) / \sqrt{2}\kappa(0) $.
Finally, the lower critical field is estimated using the formula $H_{\rm c1}(0)H_{\rm c2}(0) = H_{\rm c}(0)^{2}({\rm ln}\kappa(0) + 0.08)$.
As seen in Table \ref{tab:parameters}, we find no clear difference in the superconducting parameter, $\xi (0), \lambda (0)$, or $\kappa (0)$ for the \agyI\ or \agyII\ \AC .

We consider two possibilities for the relationship between magnetism and superconductivity:
 (i) The superconductivity mechanisms are different between \agyI\ and \agyII , and the magnetism stabilizes the superconductivity for the latter. (ii) The superconductivity mechanism is common between them and the magnetism destabilizes the superconductivity for the latter.
 At the present stage, it is unclear which of those possibilities is probable: we need further investigations.
\begin{table}[t]
\begin{center}
\caption{Superconducting parameters of the \agyI\ and \agyII\ \ACs . The underlined quantities were measured, and the others were calculated using theoretical relations described in the text.}
\label{tab:parameters}
\begin{tabular*}{\columnwidth}[b]{@{\extracolsep{\fill}}lll}
\hline
Parameter      &AGY$({\rm I})$&AGY$({\rm II})$\\
\hline
$T_{\rm c}$ $({\rm K})$&$\underline{0.68}$&$\underline{0.36}$\\
$H_{\rm c2}(0)$ $({\rm kOe})$&$\underline{3.18}$&$\underline{1.83}$\\
$H_{\rm c2}^{\rm orb}(0)$ $({\rm kOe})$&$2.76$&$1.65$\\
$H_{\rm c}(0)$ $({\rm Oe})$&$78.7$&$50.0$\\
$H_{\rm c1}(0)$ $({\rm Oe})$&$6.7$&$4.5$\\
$\xi (0)$ $({\rm nm})$&$32.2$&$42.4$\\
$\lambda (0)$ $({\rm nm})$&$9.2 \times 10^{2}$&$11.0 \times 10^{2}$\\
$\kappa (0)$&$28.6$&$25.9$\\
\hline
\end{tabular*}
\end{center}
\end{table}

In summary, we synthesized two Tsai-type 1/1 \ACs\ with a nominal composition, \agyIN\ [\agyI ] and \agyIIN [\agyII ].
By low-temperature experiments on their electrical resistivity, magnetization, ac magnetic susceptibility, and specific heat, we showed that they are the first superconductors among Tsai-type \QCs\ and \ACs . 
To account for the different magnetic properties of \agyI\ and \agyII\ \ACs ,
we proposed a model in which Yb ion located at the Tsai-type cluster center, which exists only in the \agyII\ \AC , is magnetic while the other Yb ions located at the vertex of the icosahedron are nonmagnetic. 
We further discussed the possible effect of the cluster-center magnetic Yb ion on superconductivity.
We hope that the present study stimulates further search for superconducting \QC\ and \AC , and research on the relationship between magnetism and superconductivity as well.

\section*{Acknowledgments}
The authors thank S. Watanabe and K. Miyake for valuable discussions. This work was partially supported by Grants-in-Aid for Scientific Research from JSPS, KAKENHI (Nos. 24654102, 25610094, and 26610100). K.D. also thanks the Yamada Science Foundation for financial support.

\bibliography{17006} 


%
\end{document}